\documentclass{aa}
\usepackage{graphicx}
\usepackage{txfonts}
\usepackage{natbib}
\bibpunct{(}{)}{;}{a}{}{,}    
\newcommand{\beq}{\begin{equation}}
\newcommand{\eeq}{\end{equation}}
\newcommand{\bea}{\begin{eqnarray}}
\newcommand{\eea}{\end{eqnarray}}

\newcommand{\subscr}[1]{_\mathrm{#1}}

\newcommand{\url}[1]{{\tt #1}}

\newcommand{\gsim}{\raisebox{-0.6ex}{$\stackrel{{\displaystyle>}}{\sim}$}}
\def\gapp{\lower 3pt\hbox{${\buildrel > \over \sim}$}\ }
\def\lapp{\lower 3pt\hbox{${\buildrel < \over \sim}$}\ }
\def\Mjup{{\rm M\subscr{Jup}}}

\begin{document}
\title{Disk eccentricity and embedded planets}
\author{
Wilhelm Kley
\and
Gerben Dirksen
}
\offprints{W. Kley,\\ \email{kley@tat.physik.uni-tuebingen.de}}
\institute{
     Institut f\"ur Astronomie \& Astrophysik, 
     Abt. Computational Physics,
     Universit\"at T\"ubingen,
     Auf der Morgenstelle 10, D-72076 T\"ubingen, Germany
}
\date{Received 26 July 2005 / Accepted 11 October 2005}
\abstract{} 
{We investigate the response of an accretion disk to the
presence of a perturbing protoplanet embedded in the disk through time
dependent hydrodynamical simulations.}  
{The disk is treated as a two-dimensional viscous fluid and the planet
is kept on a fixed orbit. We run a set of simulations
varying the planet mass, and the viscosity and temperature of the
disk.  All runs are followed until they reach a quasi-equilibrium
state.}
{We find that for planetary masses above a certain minimum mass,
already $3 M_{Jup}$ for a viscosity of $\nu = 10^{-5}$, the disk
makes a transition from a nearly circular state into an
eccentric state.  Increasing the planetary mass leads to a
saturation of disk eccentricity with a maximum value of around 0.25.
The transition to the eccentric state is driven by the excitation of
an $m=2$ spiral wave at the outer 1:3 Lindblad resonance.  The effect
occurs only if the planetary masses are large enough to clear a
sufficiently wide and deep gap to reduce the damping effect of the
outer 1:2 Lindblad resonance.  An increase in viscosity and
temperature in the disk, which both tend to close the gap, have an
adverse influence on the disk eccentricity.}
{In the eccentric state the mass accretion
rate onto the planet is greatly enhanced, an effect that may ease the
formation of massive planets beyond about 5 $M_{Jup}$ that are
otherwise difficult to reach.}
\maketitle
\markboth
{Kley \& Dirksen: Disk eccentricity and embedded planets}
{Kley \& Dirksen: Disk eccentricity and embedded planets}
\section{Introduction}
\label{sec:introduction}
In the early stages of their formation protoplanets are still
embedded in the disk from which they form. Not only will the protoplanet accrete
material from the disk and increase its mass,
it will also interact gravitationally with it.
Planet-disk interaction is an important aspect of planet formation
because it leads to a change in the planetary orbital elements.
Already before the discovery of extrasolar planets 
the interaction of an embedded object with a disk
has been studied for small perturber masses by linear analysis
\citep[eg.][]{1980ApJ...241..425G, 1984ApJ...285..818P, 1986Icar...67..164W},
and in more recent years also for massive planets through detailed
numerical simulations in two and three dimensions 
\citep[eg.][]{1999ApJ...514..344B,1999MNRAS.303..696K,2001ApJ...547..457K,
2002A&A...385..647D,2003MNRAS.341..213B}. 
In all these simulations the planet has been held fixed on a circular orbit and
its influence onto the disk has been analyzed. The back reaction of the disk
in terms of migration rate 
or eccentricity change can be calculated by summing over the force contribution
of each disk element.

The full evolution of a single planet embedded in a disk has been
followed for example by \citet{2000MNRAS.318...18N}.  In a later
study by \citet{2001A&A...366..263P} numerical simulations have been
performed for a range of planetary masses with emphasis on the
eccentricity evolution of the planets.  It has been found that massive
planets create an eccentric disturbance in the outer disk which in
turn may back-react on the planet and increase its eccentricity.
However, only for planets larger than 10-20 Jupiter masses a visible
increase up to $e=0.20$ has been found. However, these values are
significantly below the observed eccentricities for extrasolar planets
which average at about $e=0.3-0.4$ for planetary masses between 1 and
10 $M_{Jup}$. Also for smaller planet masses an average
eccentricity of about $e = 0.3$ is observed. This may be due to
planet-planet interactions, but these interactions are more effective
with increasing planet mass.
For a recent overview of planetary properties see
\citet{2005PThPS.158...24M} and
the {\it Extrasolar Planets Encyclopedia} ({\tt http://www.obspm.fr/encycl/encycl.html})
maintained by J.~Schneider. 
The distribution of eccentricities does not show a strong dependence
on $m \sin{i}$ nor on the distance from the central star.

As one possible scenario to explain the origin of the observed high eccentricities
the aforementioned interaction of a planet with the protoplanetary disk
has been suggested.
In particular, \citet{2003ApJ...585.1024G,
2004ApJ...606L..77S} estimate that Lindblad resonances may lead to eccentricity
growth under reasonable assumptions. Numerical simulations tend to show the
opposite, for Jupiter mass planets the eccentricity is typically damped
on short time scales $\approx 100$ orbits, only for
massive planets at least transient growth has been seen
\citep{2000MNRAS.318...18N}.
This last result may be related to the back reaction of an eccentric disk
onto the planet, where the disk's eccentricity has been induced by the presence
of the massive planet.
Additionally, the mass of the embedded planet has also profound consequences for the
mass accretion rate onto it, i.e. its growth-time.
As the induced gap in the disk becomes wider and deeper upon increasing $M_{p}$
the accretion rate diminishes and essentially limits the growth for masses
beyond 5 $M_{Jup}$ \citep{1999ApJ...514..344B, 1999ApJ...526.1001L}.
However, those calculations covered only a couple of hundred orbits of 
the planet which is much smaller than the viscous time scale.
Consequently, no equilibrium structure has been reached.

Here we follow this line of thought and investigate the influence a massive
embedded planet has on the structure of the ambient protoplanetary disk.
We use a hydrodynamical description to follow the evolution of the disk, where
the planet is fixed on a circular or a slightly eccentric orbit.
All simulations are run until a quasi-stationary equilibrium has been reached
and overall values of mass and energy in the computational domain remain
unchanged.
We vary the mass of the planet, the temperature and the disk viscosity, and analyze their
influence on the structure of the disk, in particular on its eccentricity.
Indeed, we find that (for a given viscosity) there appears to be clear
transition in the disk from an circular state into an eccentric state.

We analyze the magnitude of the induced disk eccentricity and estimate its 
influence on the accretion rate of the planet.
In particular, we find that for sufficiently massive planets the disk becomes
eccentric, where the critical minimum mass depends on the value of the
viscosity coefficient.

For a viscosity of $\alpha = 4 \times 10^{-3}$, a reasonable
value for protoplanetary disks, the disk becomes eccentric already
for planets of 3 Jupiter masses. At the same time the mass accretion rate
onto the planet increases strongly for an eccentric disk.

In the next section we describe our model assumptions, in section 3 we present
our results followed by theoretical analysis and conclusions.

\section{The Standard Hydrodynamical Model}
\label{sec:hydro-model}
The models presented here are calculated 
basically in the same manner as those described previously in
\citet{1998A&A...338L..37K, 1999MNRAS.303..696K}.
The reader is referred to those papers
for details on the computational aspects of this type of simulations.
Other similar models, following explicitly the motion of single
planets in disks, have been presented by
\citet{2000MNRAS.318...18N}, \citet{2000ApJ...540.1091B}.
 
We use cylindrical coordinates ($r, \varphi, z$) and 
consider a vertically averaged, infinitesimally thin disk located
at $z=0$. The origin of the coordinate system, which is co-rotating
with the planet, is either
at the position of the star or in the combined
center of mass of star and planet.
Since in the first case the coordinate system is accelerated and
rotating, care has to be taken to include also the indirect terms of
the acceleration \citep{1998A&A...338L..37K}.
The basic hydrodynamic equations (mass and momentum conservation)
describing the time evolution of such a viscous
two-dimensional disk with embedded planets have been stated frequently and are
not repeated here \citep[see][]{1999MNRAS.303..696K}.

In the present study we restrict ourselves to the situation where the
embedded planet is on a fixed orbit, i.e. the gravitational
back reaction of the disk on the planet is not taken into account.
\subsection{Initial Setup}
The two-dimensional ($r - \varphi$) computational domain consists of a
complete ring of the protoplanetary disk.
The radial extent of the computational domain
(ranging from $r\subscr{min}$ to $r\subscr{max}$) is taken such that there is
enough space on both sides of the planet, although, as we shall see later,
the effect we are analyzing appears to occur only in the outer disk.  
Typically, we assume $r\subscr{min}=0.40$ and for $r\subscr{max}$ we take
two different values 2.5 and 4.0, in units where the planet is located at $r=1$.
In the azimuthal direction for a complete annulus we have $\varphi\subscr{min} =0$ and
$\varphi\subscr{max} = 2 \pi$.

The initial hydrodynamic structure of the disk (density, temperature, velocity)
is axisymmetric with respect to the location of the star.
The surface density is constant ($\Sigma=1$ in
dimensionless units) over the entire domain with no initial gap.
To make sure that only little disturbances or numerical artifacts
arise upon immersion of the planet, its 
mass will be slowly turned on from zero to the final required mass (eg. 5 Jupiter masses)
over a time span of typically 50 orbital periods.
The initial velocity is pure Keplerian rotation ($u_r=0,
u_\varphi = \Omega_K r = (G M_*/r)^{1/2}$), and 
the temperature stratification is always given by
$T(r) \propto r^{-1}$ which follows from an assumed
constant vertical height $H/r$.
For these isothermal models the temperature profile is left unchanged
at its initial state throughout the computations.

For our standard model we use a constant kinematic viscosity coefficient $\nu$
but present additionally a sequence of $\alpha$-disk models.
\subsection{Boundary conditions}
To ensure a most uniform environment for all models and
minimize disturbances (wave reflections) from the outer
boundary we impose at $r\subscr{min}$ and $r\subscr{max}$ damping boundary
conditions where the density and both velocity components
are relaxed towards their initial values as
\begin{equation}
   \frac{d X}{d t}  =  - \frac{ X  - X(t=0)}{\tau\subscr{damp}}  \, R(r)^2  
\end{equation}
where $X \in \{\Sigma, u_r, u_\varphi\}$,
$\tau\subscr{damp} = 1/\Omega\subscr{K}(r\subscr{boundary})$ and $R(r)$
is a dimensionless linear ramp-function rising from 0 to 1 from
$r\subscr{damp}$ to $r\subscr{boundary}$. Here, 
$r\subscr{boundary}$ is either $r\subscr{min}$ or $r\subscr{max}$, depending
which edge of the disk is considered.
The initial radial velocity vanishes, and the boundary conditions
ensure that no mass flows through the radial boundaries at 
$r\subscr{min}$ and $r\subscr{max}$. However, the total mass in the system may 
nevertheless vary due to the applied damping.
In the azimuthal direction, periodic boundary conditions for all
variables are imposed.

These specific boundary conditions allow upon a long term evolution
for a well defined quasi-stationary state
if there is no back-reaction of the disk on the orbital elements
of the planet.
\subsection{Model parameters}
The computational domain is covered by 128 $\times$ 384 ($N_r \times N_\varphi$)
grid cells for the smaller models $[0.4,2.5]$ and 
$200 \times 384$ for the larger $[0.4,4.0]$ ones.
The grid is spaced equidistant in both radius and azimuth.
The inner radius beyond which the damping procedure defined above gradually sets in
is given by $r\subscr{damp} = 0.5$, the outer damping radius is given by 
$R\subscr{damp} = 0.84 r\subscr{max}$. 
The star has a mass of 1 $M_\odot$, and the
mass of the planet in the different models
ranges from one to five Jupiter masses. 
The planet is held on a fixed circular orbit.

For the viscosity a value of $\nu = 1.0 \cdot 10^{-5}$ (in units of
$\Omega_p r_p^2$) is used for our
standard models, which is equivalent to a value of $\alpha = 0.004$
for the standard $H/r = 0.05$.  This is a typical value for the
effective viscosity in a protoplanetary disk.

To achieve a more detailed calculation of the observed phenomena we
refined some calculations to the higher resolution of 260 x 760 ($N_r
\times N_\varphi$) by interpolating the data from coarser
calculations. As the relaxation time for the system is very long ($>$
1000 orbits) it would be too time-consuming to complete the whole
calculation on the high-resolution grid. These higher resolution
simulations yield identical results.
To study the influence of physical parameters such as viscosity and pressure, we vary
$\nu$ and $H/r$ in some models.
In addition, we analyze the influence of several numerical parameters on the results.

\subsection{A few remarks on numerical issues}
We use two different codes for our calculations, {\tt RH2D} and {\tt
NIRVANA}.  The numerical method used in both codes is a staggered
mesh, spatially second order finite difference method based, where
advection is based on the monotonic transport algorithm
\citep{1977JCoPh..23..276V}.  Due to operator-splitting the codes are
semi-second order in time.  The computational details of {\tt RH2D}
which can be used in different coordinate systems have been described
in general in \citet{1989A&A...208...98K}, and specifically for planet
calculations in \citet{1999MNRAS.303..696K}.  The details of the {\tt
NIRVANA} code have been described in \citet{1998CoPhC.109..111Z}.

The use of a rotating coordinate system requires special treatment 
of the Coriolis terms to ensure angular momentum conservation
\citep{1998A&A...338L..37K}.
Especially for the long-term calculations presented here, this is an 
important issue.

In calculating the gravitational potential of the planet we use a
smoothed potential of the form
\beq
    \Phi_P  =  - \frac{ G M\subscr{p}}{\sqrt{s^2 + \epsilon^2}}
\eeq
where $s$ is the distance from the planet.
For the smoothing length of the potential we choose $\epsilon = 0.4 R\subscr{Hill}$.

The viscous terms, including all necessary tensor components,
are treated explicitly.
To ensure stability in the gap region with very strong gradients in the
density an artificial bulk viscosity has been added, with a
coefficient $C\subscr{art} = 1.0$. For the detailed formulation
of the viscosity related issues and tests see \citet{1999MNRAS.303..696K}.

As the mass ratio $M\subscr{p}/M_*$ of the planet can be very large
we have found it preferable to work with a density floor, where the density
cannot fall below a specified minimum value $\Sigma\subscr{min}$.
For our purpose we use a value of $\Sigma\subscr{min} = 10^{-8}$ in dimensionless values,
where the initial density is of ${\cal{O}}(1)$.

\section{The dual-state disk}
\begin{figure}[ht]
\begin{center}
\resizebox{0.98\linewidth}{!}{%
\includegraphics{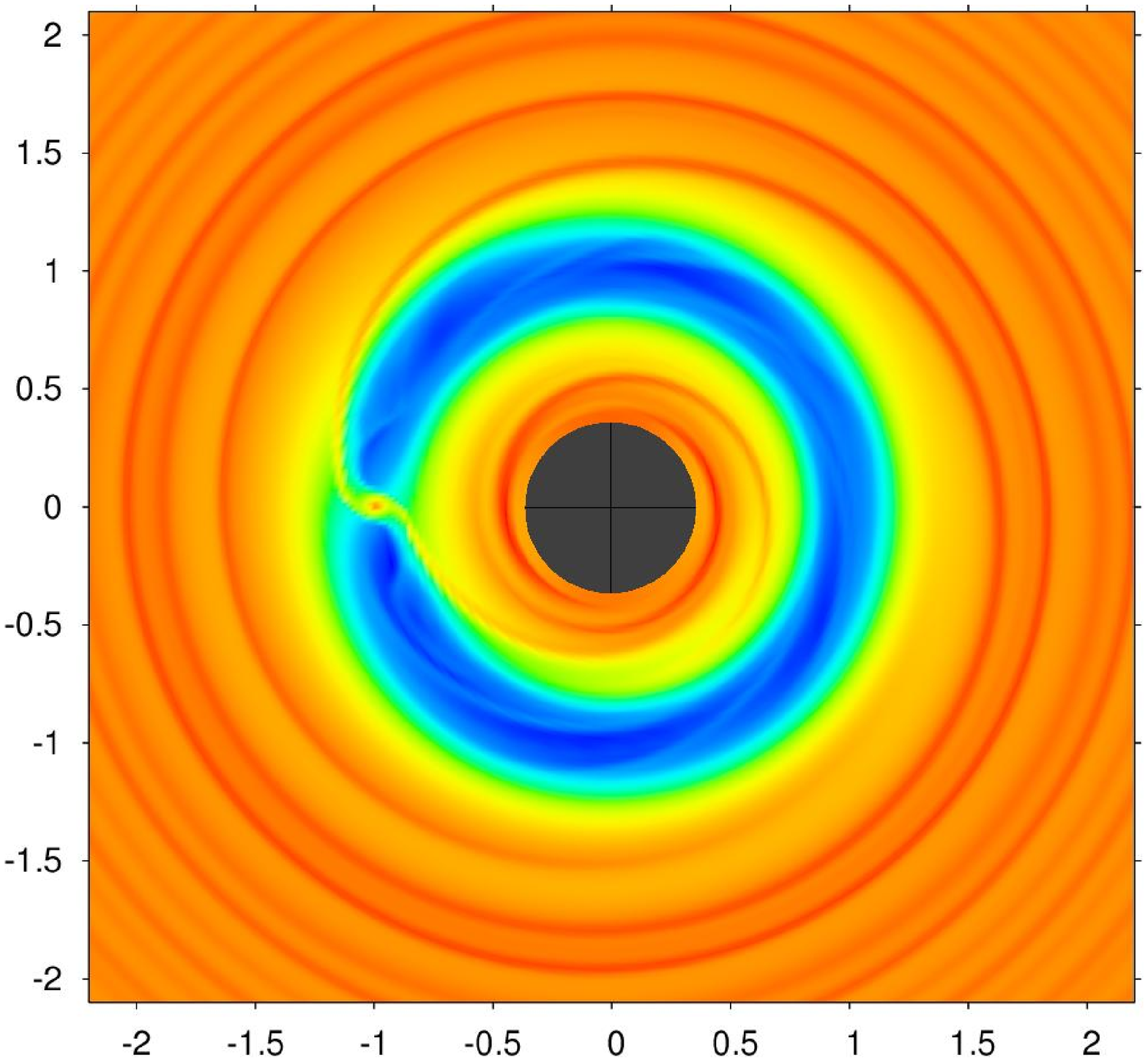}}
\resizebox{0.98\linewidth}{!}{%
\includegraphics{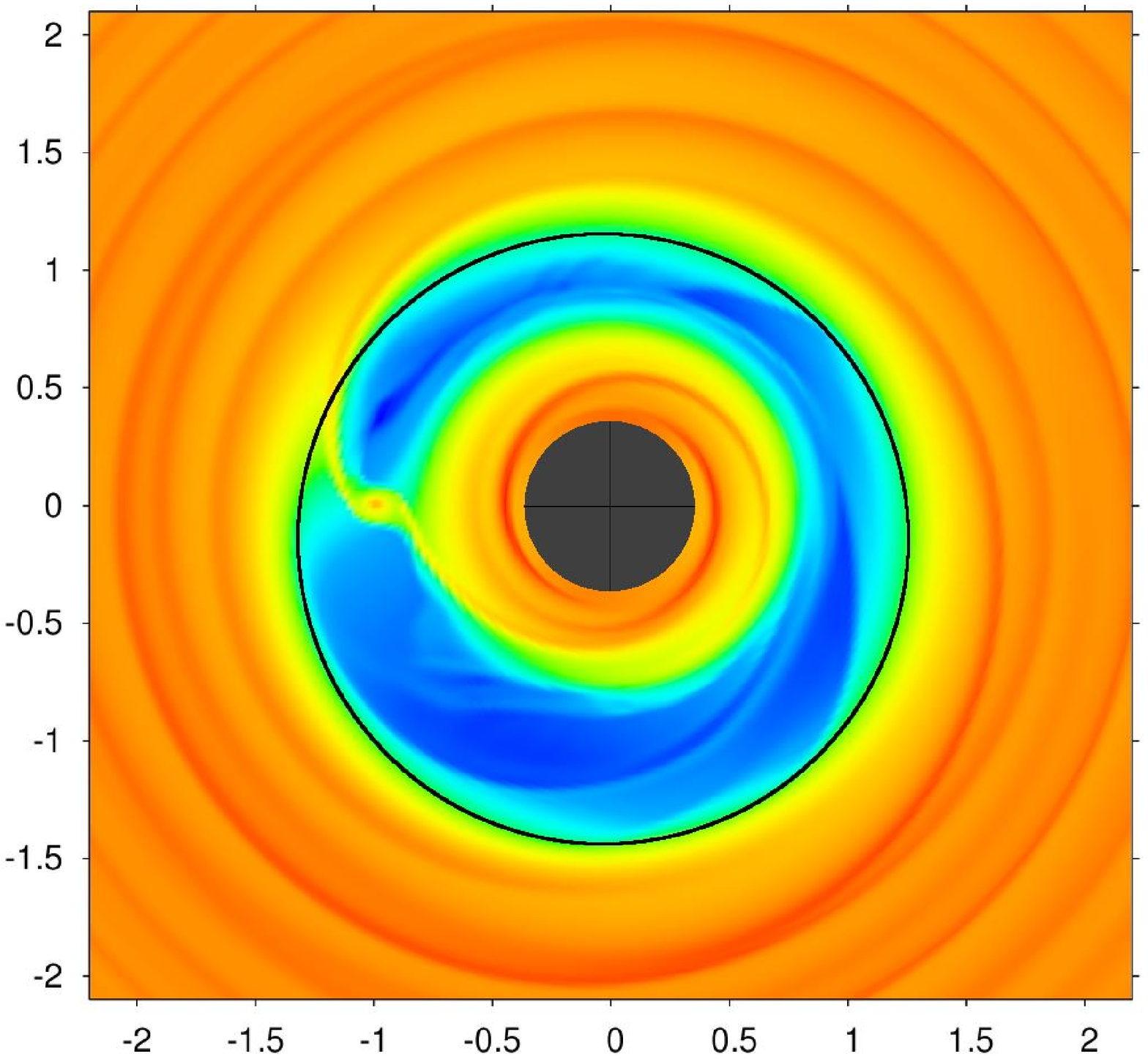}}
\end{center}
  \caption{ Logarithmic plots of the surface density $\Sigma$ for the
  relaxed state after 2000 orbits for two different masses of the
  planet which is located at $r=1.0$ in dimensionless units.  {\bf
  Top}) $q = 3.0 \, 10^{-3}$ and {\bf Bottom}) $q = 5.0 \, 10^{-3}$
  calculated with NIRVANA.  The inner disk stays circular in
  both cases but the outer disk only in the lower mass case. For $q =
  5.0 \, 10^{-3}$ it becomes clearly eccentric with some visible fine
  structure in the gap.  For illustration, the drawn ellipse (solid
  line in the lower plot) has one focus at the stellar location and an
  eccentricity of 0.20.  }
   \label{fig:equil}
\end{figure}

We first consider our standard model as described above
using a planetary mass ranging from 1 to 5 $M_{Jup}$, i.e.
a mass ratio of $q = 10^{-3}$ to $5 \cdot 10^{-3}$.
The other physical parameters are identical for all models.
Due to the nature of the damped boundary conditions and a non-zero
physical viscosity we might expect after a sufficiently long evolution
time a convergence towards an equilibrium state where the density
structure and the total amount of mass in the disk remain 
constant in time, at least in the co-rotating frame.
Indeed, for small planetary masses,
$M_{p} < 3 M_{Jup}$ we find a circular stationary state
which displays the typical features of embedded planets in disks: a deep,
circular depression of density at the location of the planet (the gap), spiral
arms in the inner and outer disk.
This state is shown in the top graph of Fig.~\ref{fig:equil},
which shows the surface density of
the obtained equilibrium state at an evolutionary time of $t=2000$ orbits. 

However, if the planetary mass reaches $M_{p} \geq 3 M_{Jup}$ we
surprisingly do not reach a stationary equilibrium state
anymore. Instead we find after a very long time ($> 1000$ orbits) a
new periodic state which has approximately the same period as the
orbital period of the planet.  In this state the disk is clearly
eccentric with an extremely slow precession rate such that the
eccentric pattern appears to be nearly stationary in the inertial
frame.  This eccentric quasi-equilibrium state for $M\subscr{p} = 5
M_{Jup}$ is shown in the bottom graph of Fig.~\ref{fig:equil}.

\subsection{The eccentric disk}
\begin{figure}[ht]
\begin{center}
\resizebox{0.98\linewidth}{!}{%
\includegraphics{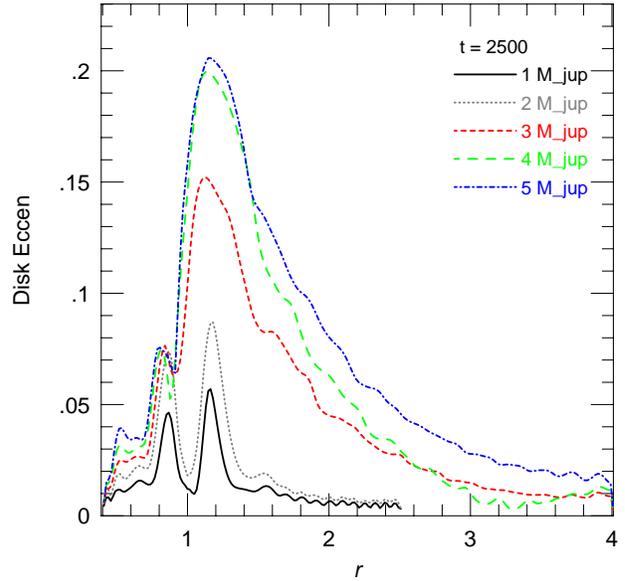}}
\end{center}
  \caption{Disk eccentricity as a function of radius for the several models with
 $q = 0.001$ up to $q = 0.005$ at $t=2500$ orbits, for the $q=0.003$ model at
  $t=3850$.
 For the two lower curves $q = 0.001$ and $q = 0.002$,
 the outer edge of the computational domain lies at $r_{max} = 2.5$.
    }
   \label{fig:ecc1-mp}
\end{figure}

A measure of the eccentricity of the disk is calculated as follows:
For a ring at radius $r_i$ we calculate the eccentricity for every
cell in the ring from the velocity and position vector of that cell by
assuming the fluid element is a particle moving freely in the central
potential of the star, feeling no pressure forces. The average over
all cells in the ring is then defined as the eccentricity of the disk
at that radius $r_i$.  This value is plotted for different masses in
Fig.~\ref{fig:ecc1-mp} at the evolutionary time of $t=2500$, only for
$M_p = 3 M_{Jup}$ at $t=3850$ orbits.
 
\begin{figure}[ht]
\begin{center}
\resizebox{0.98\linewidth}{!}{%
\includegraphics{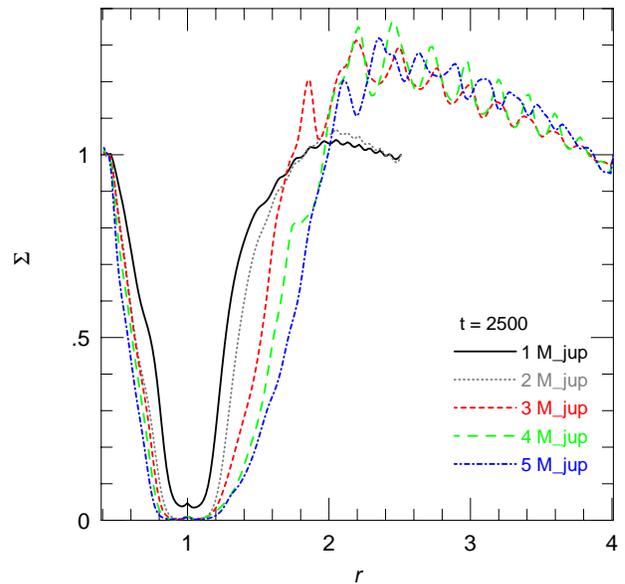}}
\end{center}
  \caption{Azimuthally averaged radial profiles of the surface density
  for different planet masses,
 for the same models and times as in Fig.~\ref{fig:ecc1-mp}.
 The width of the gap increases with planetary mass.
    }
   \label{fig:sig1-mp}
\end{figure}

For planetary masses below around $M_p \approx 3 M_{Jup}$, the maximum
eccentricity of the disk is about 0.10, and is strongly peaked at $r
\approx 1.2$.  For the larger planetary masses the eccentricity of the
disk nearly doubles and reaches 0.22 for $M_p = 5 M_{Jup}$.  In
addition, a much larger region of the disk has become eccentric, which
has been seen clearly already in the surface density distribution in
Fig.~\ref{fig:equil}, bottom, where the ellipse indicates an
eccentricity of 0.20 with one focus at the stellar position.  The
precession rate $\dot{\varpi}$ of the eccentric disk is very small and
typically prograde.  From our longest runs (over several thousand
orbits) we estimate $\dot{\varpi} \approx 10\deg / 1000$ Orbits.  In
Fig.~\ref{fig:ecc1-mp} the curves for the lower planet masses end at
$r = 2.5$ because this is the outer boundary for those low mass
models.

In Fig.~\ref{fig:sig1-mp} the azimuthally averaged density profile is
plotted for different planetary masses for the same models
as in Fig.~\ref{fig:ecc1-mp}. Clearly the gap width increases for the larger
planet mass, as expected due to the stronger gravitational torques.
For the lowest mass
$q=0.001$ model (solid line) the gap is not completely cleared.
\begin{figure}[ht]
\begin{center}
\resizebox{0.98\linewidth}{!}{%
\includegraphics{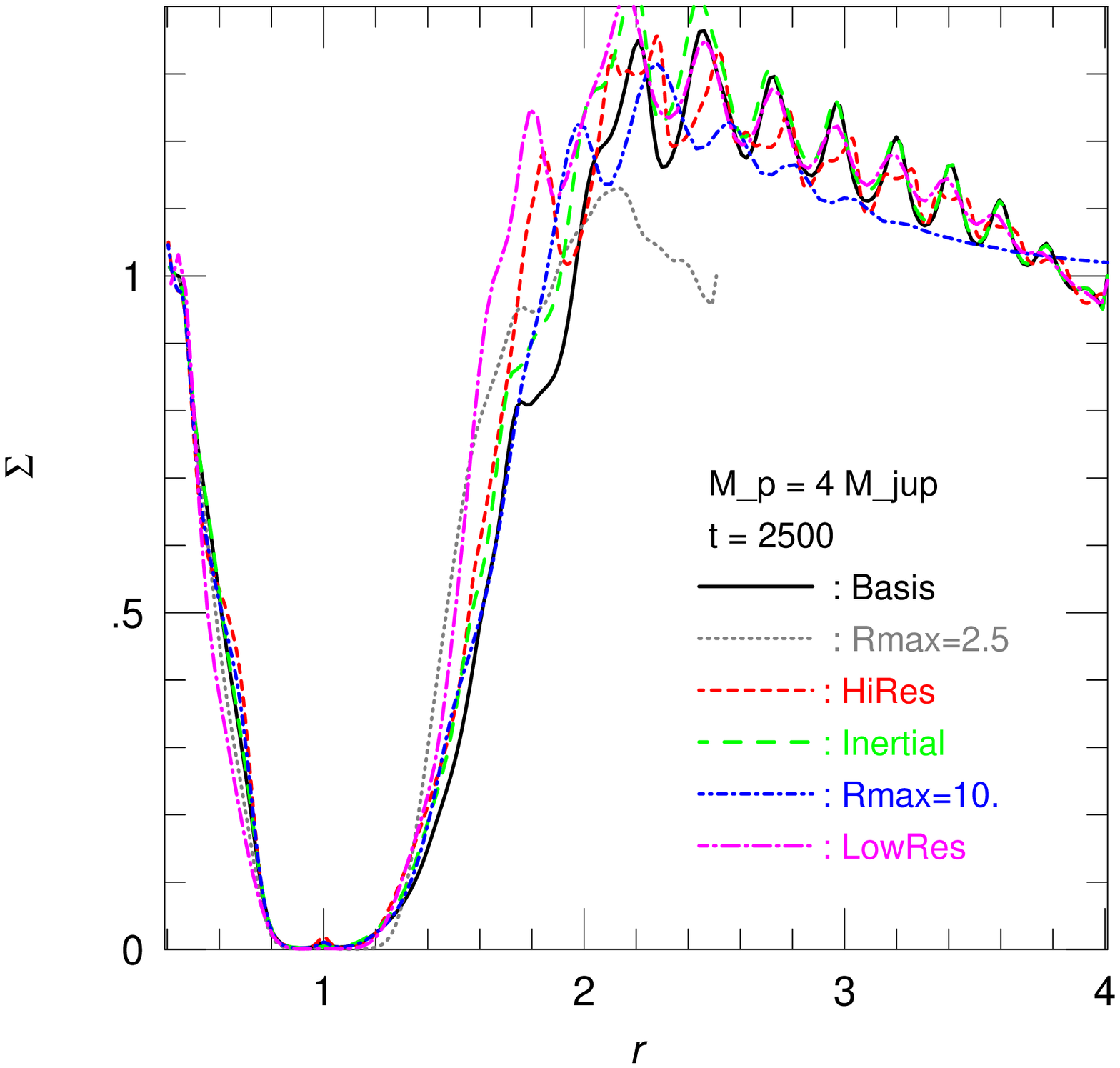}}
\resizebox{0.98\linewidth}{!}{%
\includegraphics{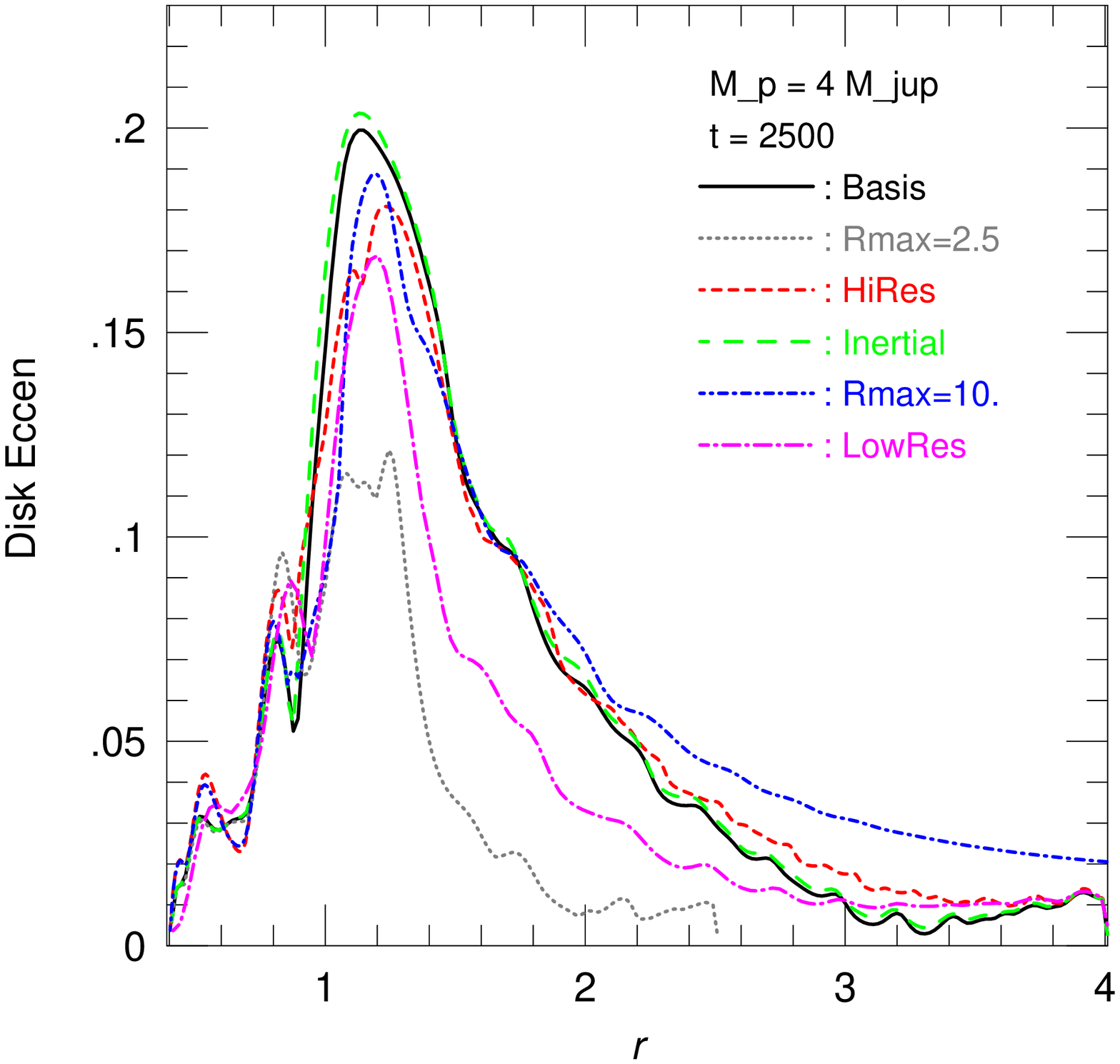}}
\end{center}
  \caption{
  Surface density and eccentricity profile for models using
   $q = 0.004$ at a time of 2500  orbits,
  the high resolution (model short dashed line) at 1750 orbits.
   Plotted are results for different models varying the numerical
  setup.
    }
   \label{fig:sigecc-num}
\end{figure}
\subsection{Dependencies on numerical parameters}
The threshold mass where the transition from circular to eccentric occurs
apparently depends on the width and shape of the gap, and parameters that
will change the gap structure will also change this threshold mass. 
Before we analyze physical influences we display
in Fig.~\ref{fig:sigecc-num} the surface density profile
and the disk eccentricity for models
using different numerical parameters but all with same physical setup
for $q = 0.004$, and
at the same evolutionary time of 2500 orbits (the high resolution
model at $t=1750$ orbits).

The solid line refers to the basic reference model (as in
Fig.~\ref{fig:sig1-mp}, $4 M_{Jup}$ model).  We first find that the
mass value where the transition occurs may depend on the location of
the outer boundary $r\subscr{max}$. If the stand-off distance of the
planet to the outer boundary is too small the damping boundary
conditions, which tend to circularize the disk, prevent the disk from
becoming eccentric.  The simulations using a $4 M_{Jup}$ planet and a
smaller $r\subscr{max}$ clearly shows this effect.  For this mass of
the planet the disk will not anymore become eccentric for
$r\subscr{max} = 2.5$ (dotted curve).  Hence, to properly study this
effect a sufficiently large $r_{max}$ has to be chosen. An extended
domain with $r\subscr{max} = 10$ (short-dashed-dotted) does not alter
the eccentricity behavior of the disk.  The inner disk remains
circular for all planet masses because of the strong damping
introduced by the boundary condition.

A higher resolution (200 $\times$ 500, short-dashed line), and running
the model in the inertial frame (long-dashed) have no significant
influence on the density distribution and the occurrence and magnitude
of the disk eccentricity. A lower resolution model
(long-dashed-dotted) using 128 $\times$ 128 grid cells, results in a
slightly lower eccentricity due to a larger (numerical) damping.  In
addition, we have compared results with two different numerical codes
({\tt RH2D} and {\tt NIRVANA}) and again found good agreement. Hence,
we conclude that the eccentric disk state is a robust,
reproducible physical phenomenon.
\begin{figure}[ht]
\begin{center}
\resizebox{0.98\linewidth}{!}{%
\includegraphics{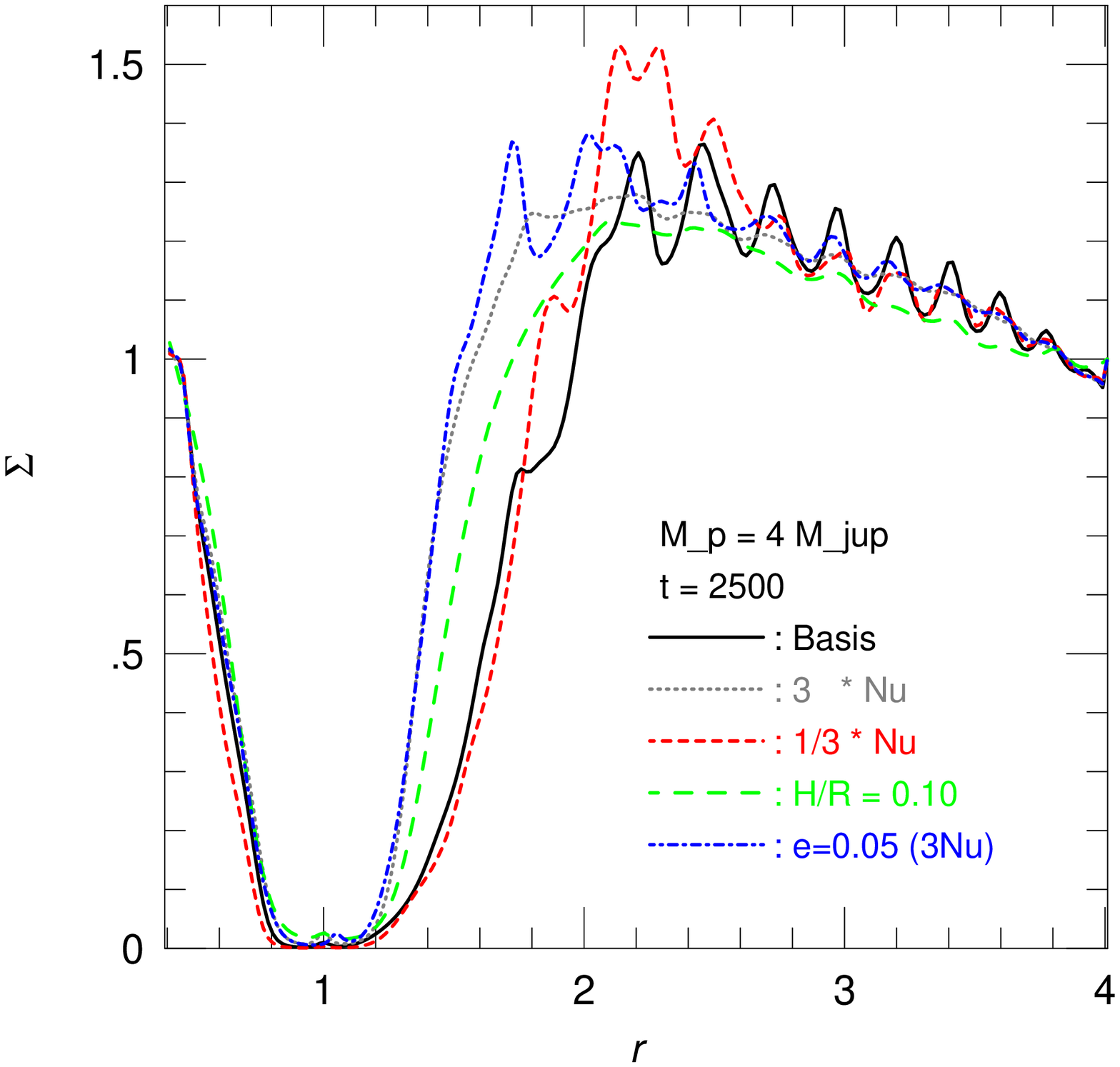}}
\resizebox{0.98\linewidth}{!}{%
\includegraphics{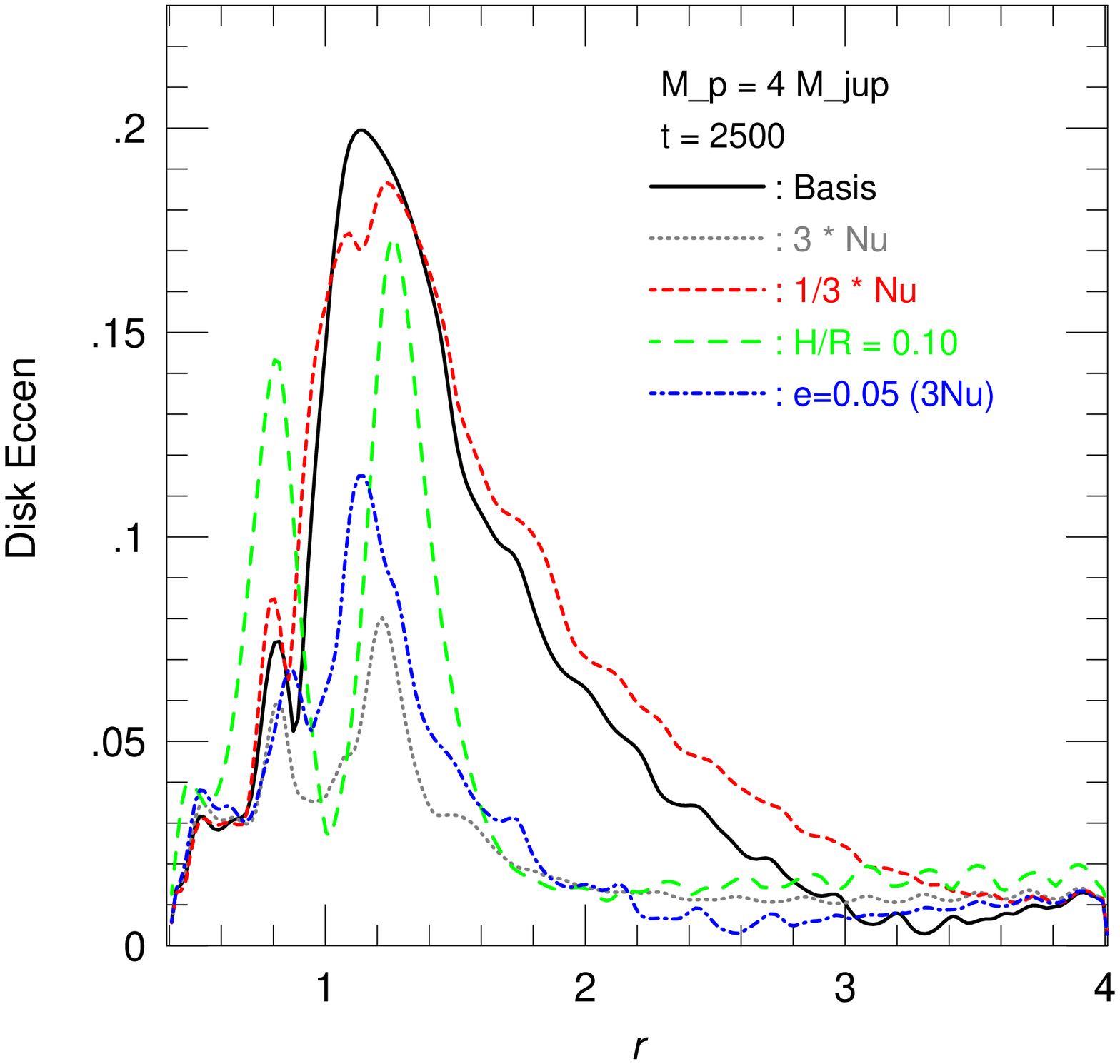}}
\end{center}
  \caption{
  Surface density and eccentricity profile for models using
   $q = 0.004$ at a time of 2500 orbits. 
   Plotted are results for different models varying the physical
  setup.
    }
   \label{fig:sigecc-phy}
\end{figure}
\subsection{Dependencies on physical parameters}
In Fig.~\ref{fig:sigecc-phy} we display the surface density profile
and the disk eccentricity for models with $q = 0.004$ using different
physical parameters.  If the dimensionless viscosity $\nu$ is enlarged
to $3 \times 10^{-5}$ (dotted line) the gap width and depth is reduced
and the disk will no longer become eccentric for the planet mass of $q
= 0.004$ (and also not for $q = 0.005$).  Similarly, an increased
$H/r$ (long-dashed line) leads also to a narrower gap and a smaller disk
eccentricity.  If, on the other hand, the viscosity is lowered by a factor
of three (short-dashed), or $H/r$ is reduced we find that the disk
reaches about the same eccentricity as before.

The last model (dashed-dotted line) refers to a planet on an eccentric orbit
with $e_p=0.05$ and a 3 times higher viscosity than the basis model.
As can be seen, the disk remains circular for these parameter.
This model demonstrates that it is not the planetary eccentricity which is
responsible for producing the disk eccentricity but that it is rather
a genuine instability. This conclusion is confirmed by a model
with $M_{p} =2  M_{Jup}$ and $e_p = 0.05$ which (for the standard viscosity)
does not produce an eccentric disk.

\subsection{The two equilibrium states for an $\alpha$ type viscosity}
\label{subsect:alpha}
To illustrate the effect under different physical conditions we
present additional simulations using a slightly different setup.
Here, we consider a planet moving inside a disk at a radius of 0.35
AU, assuming that the inner disk has been cleared already.  The outer
radius of the computational domain lies at 1.2 AU, and the inner one
at 0.25 AU. The scale height of the disk is $H/r = 0.05$, and for the
viscosity we use here as an alternative an $\alpha$-prescription, with
a constant value of $\alpha = 0.01$.  In these models we have used a
planetary eccentricity of $e_p = 0.01$ which is typically found in
models of embedded planets that follow the orbital evolution. As shown
above this value of $e_p$ has no influence on the transition to the
eccentric disk state.  The remaining setup is similar to the models
described above.  The viscosity may be on the large side of
protoplanetary disks but has (in combination with the lack of the
inner disk) the clear advantage of speeding up the simulations
considerably which allows us to reach the quasi-equilibrium states in
which global quantities such as mass, energy do not vary in time
anymore, with reasonable computational effort.  This alternative setup
has been used recently in a paper modeling the resonant system GJ 876
and it is described in more detail in
\citet{2005A&A...437..727K}. Here we describe additional results
concerning details of the eccentric disk state.
\begin{figure}[ht] 
\begin{center}
\resizebox{0.98\linewidth}{!}{%
\includegraphics{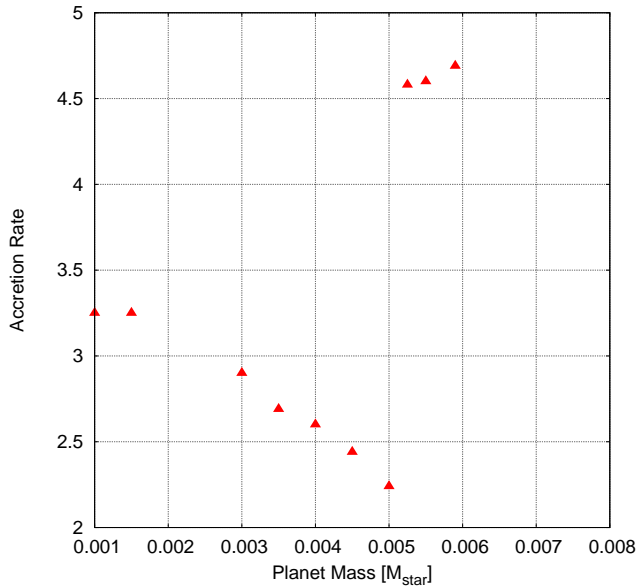}}
\end{center}
  \caption{
  The dependence of
  the accretions rate onto the planet
  (in dimensionless units) on the planetary mass
  for relaxed quasi-equilibrium configurations. 
  Results are displayed for
  models using an $\alpha = 0.01$ viscosity. 
    }
   \label{fig:mdisk-mp}
\end{figure}

For these $\alpha$-models we vary the planet-star mass ratio $q$ from
$1 \cdot 10^{-3}$ to about $7 \cdot 10^{-3}$. In all cases the models
are evolved until a quasi-stationary state has been reached.  As
already seen above for the constant viscosity case, also in this case
the disk changes its structure from circular for small planetary
masses to eccentric for large planetary masses.  Here the transition
occurs at a larger planetary mass because of the higher effective
viscosity.  

\begin{figure}[ht]
\begin{center}
\resizebox{0.98\linewidth}{!}{%
\includegraphics{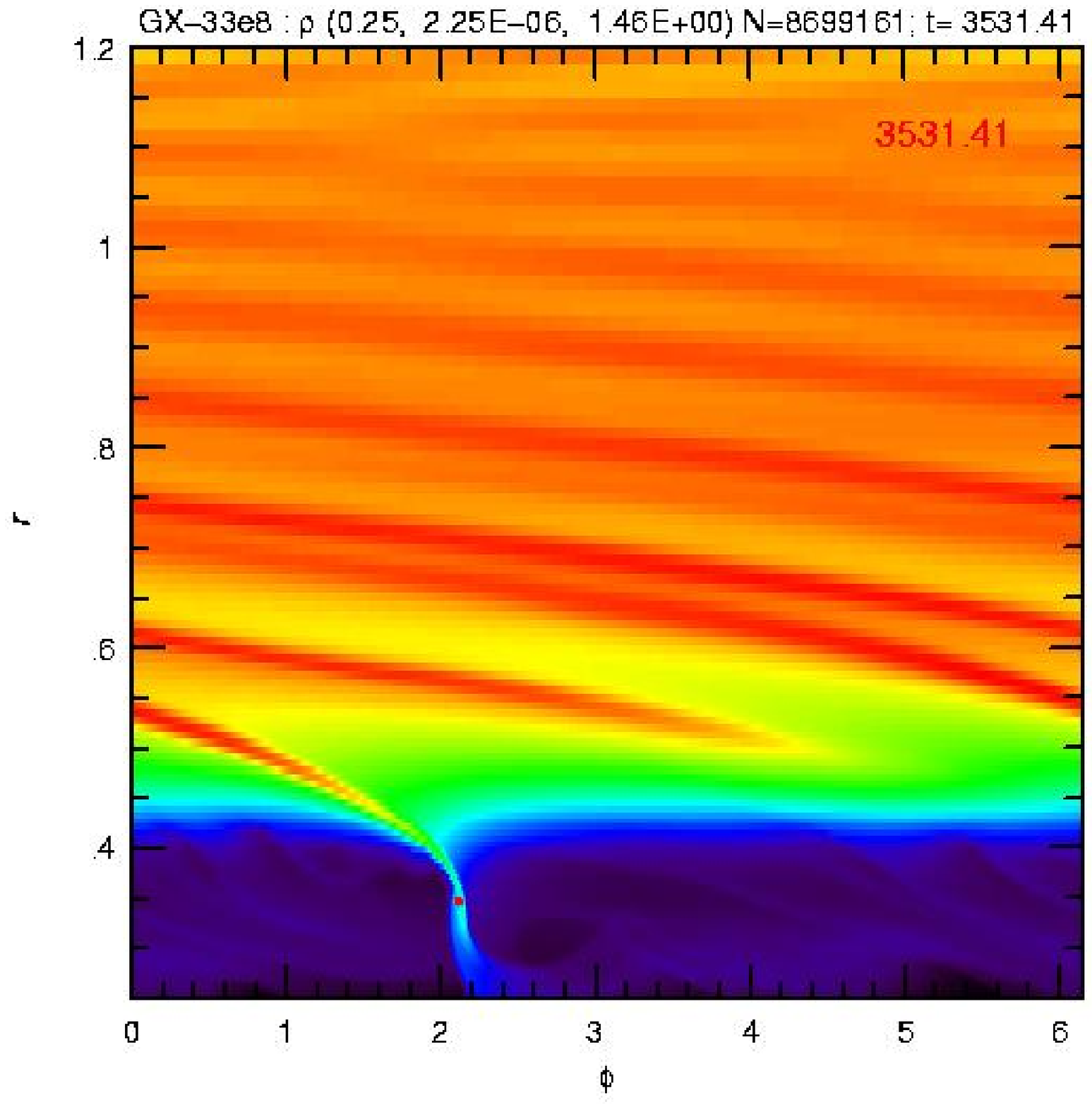}}
\resizebox{0.98\linewidth}{!}{%
\includegraphics{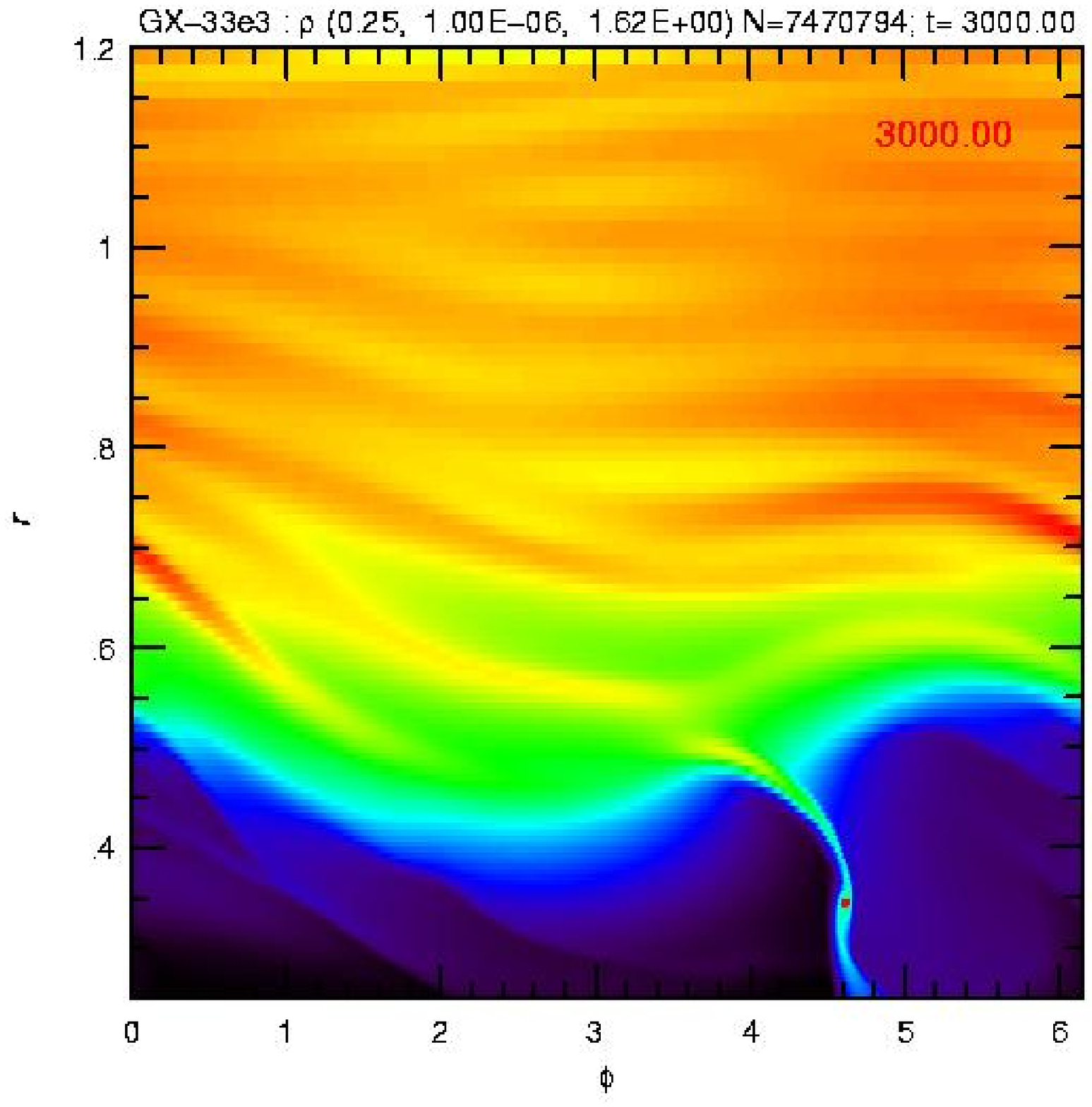}}
\end{center}
  \caption{
  Gray scale plots of the surface density $\Sigma$ for the relaxed state
   for two different planetary masses: {\bf a}) $q = 4.5 \cdot 10^{-3}$
   and {\bf b}) $q = 5.9 \cdot 10^{-3}$ calculated with RH2D.
  Due to the higher planetary mass much stronger wave-like disturbances are
  created in the density.
    }
   \label{fig:sigma2d}
\end{figure}

In Fig.~\ref{fig:mdisk-mp} we display the mass
accretion rate onto the planet as a function of the planet mass.
There is a strong jump in the magnitude of the accretion rate at a
critical planetary mass $q_{crit} \approx 5.25 \cdot 10^{-3}$,
exactly at the point where the
disk switches from circular to eccentric.
For small planetary masses $q < q_{crit}$ the mass accretion rate falls
off with increasing planetary mass, because upon increasing $M_{p}$
the stronger gravitational torques will deepen the gap and reduce the
accretion rate \citep{1999ApJ...514..344B, 1999ApJ...526.1001L}.
However, when the disk turns eccentric the gap edge periodically
approaches the planet and it may even become engulfed in the disk
material for sufficiently large eccentricity (see
Fig.~\ref{fig:sigma2d}).  Consequently, the mass accretion rate onto
the planet is strongly increased allowing for more massive planets.
 
This sudden change in the accretion rate is reminiscent of a {\it phase
transition} where the ordering parameter is given here by the
planetary mass.  Test simulations have shown that the obtained
equilibrium structure does not depend on the initial configuration
(eg. density profile, initial mass in the disk) but is solely given by
the chosen physical parameters.  As shown above the transition from
the non-eccentric state to the eccentric state, which is here a
function of only the planetary mass, depends also on the viscosity and
temperature on the disk which we have held fixed in this model
sequence.

Similarly to accretion rate the total disk mass contained in the system
also changes abruptly at the $q_{crit}$ as
a consequence of the applied the boundary conditions
at $r_{max}$. These are chosen such that the disk relaxes towards
its initial conditions at the outer boundary, eg. the value of the surface
density is fixed at that point. Upon increasing the planet mass
the gap becomes more pronounced and disk mass is pushed towards the outer
boundary increasing the density there. At the onset of the eccentric
state this relation changes abruptly.

The existence of the two equilibrium states of the disk is further
illustrated in Fig.~\ref{fig:sigma2d} where we display gray scale
plots of the surface density $\Sigma$ for the relaxed state.
for two different mass ratios ($q = 4.5$ and $5.9 \cdot
10^{-3}$) in a $r - \varphi$ representation.  While for the lower mass
case ($q = 4.5 \cdot 10^{-3}$) the disk structure remains quite
regular, the second high mass case ($q = 5.9 \cdot 10^{-3}$) shows a
strongly disturbed disk which has gained significant eccentricity ($e
= 0.2$) where also the gap edge becomes highly deformed (compare to
Fig.~\ref{fig:equil}).
\begin{figure}[ht]
\begin{center}
\resizebox{0.98\linewidth}{!}{%
\includegraphics{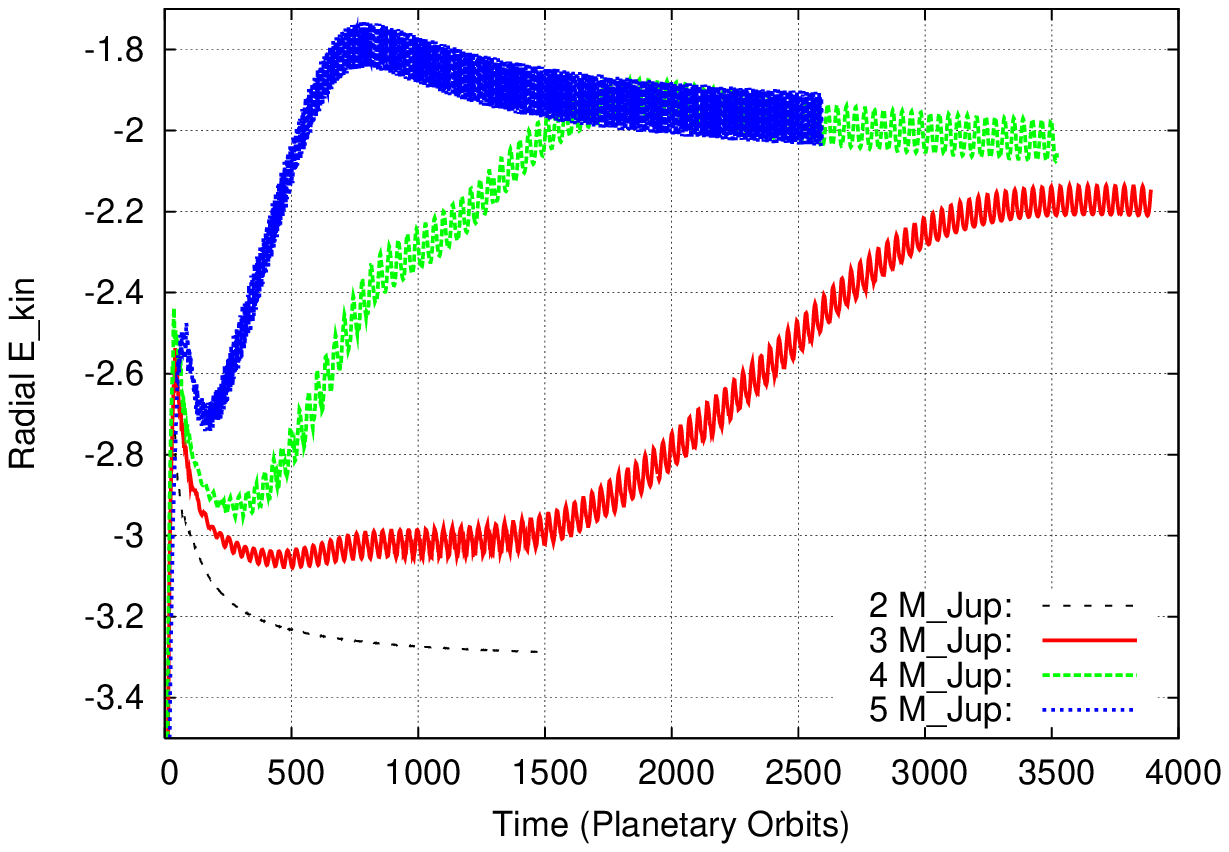}}
\resizebox{0.98\linewidth}{!}{%
\includegraphics{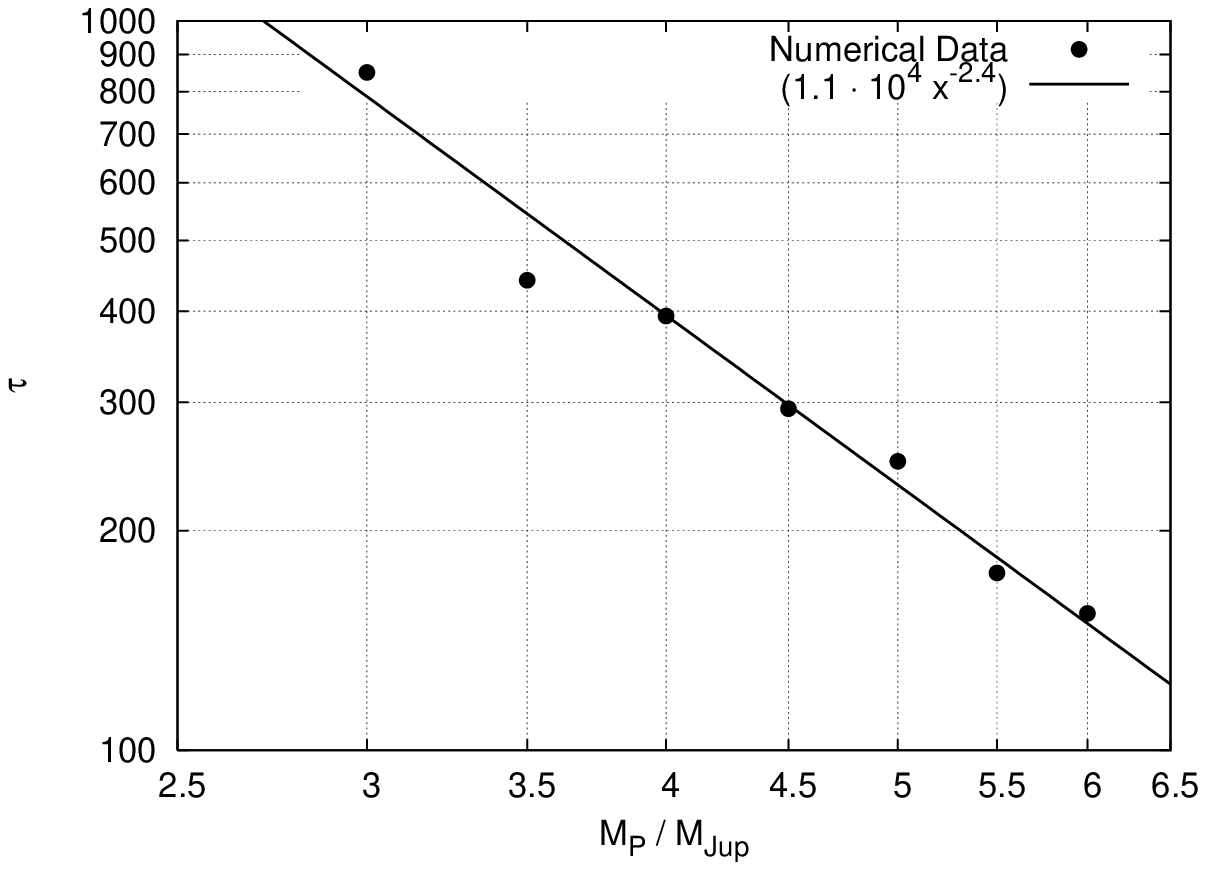}}
\end{center}
  \caption{
  {\bf a)} The time dependence of the total radial kinetic energy of the disk
  in the computational domain for four different planet masses. 
  {\bf b)} the growth rate of the eccentric disk mode as a function of 
 the planetary mass. The superimposed straight line has a slope
  of $\tau \propto {M_{p}}^{-2.4}$.
    }
   \label{fig:ekingrow}
\end{figure}

\subsection{Eccentricity growth rates}
The growth of the eccentricity of the disk depends primarily on the
mass of the planet. To measure the speed of the increase we analyze
the time dependence of the total radial kinetic energy $E_{kin,rad}$
in the models, because this is a quantity most readily available.  In
the top panel of Fig.~\ref{fig:ekingrow} we display the $E_{kin,rad}
(t)$ for four different planet masses.  For a low mass of $M_{p} =2
M_{Jup}$ no growth is visible but for larger planets the growth time
shortens upon increasing $M_{p}$.  From the growth of $E_{kin,rad}
(t)$ we estimate visually the growth-times $\tau$ as a function of
planetary mass (lower panel of Fig.~\ref{fig:ekingrow}).  Clearly, for
more massive planets the disk will turn eccentric much faster. From
the plot we may estimate a growth rate $\gamma = 1/\tau \propto
{M_{p}}^{2.4}$, a relation which is indicated by the additional
straight line in the graph. This dependence on planetary mass is
somewhat stronger than that estimated on theoretical grounds
\citep{2001A&A...366..263P}.

\begin{figure}[ht]
\begin{center}
\rotatebox{270}{
\resizebox{0.98\linewidth}{!}{%
\includegraphics{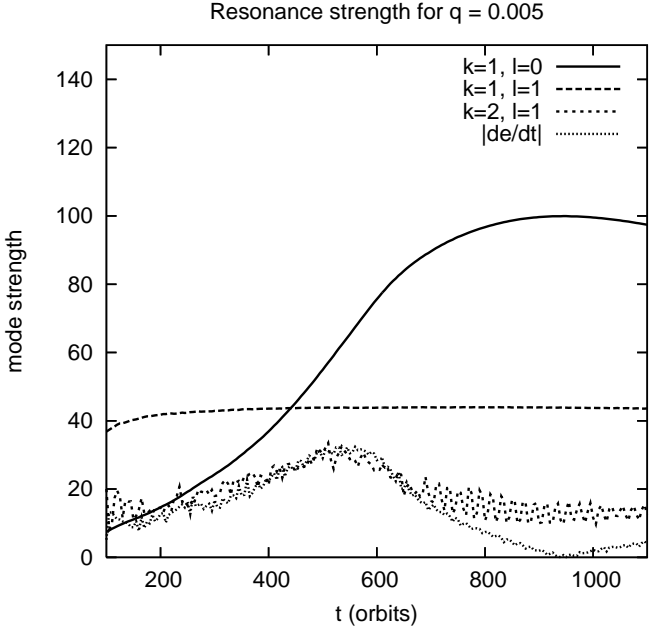}}}
\end{center}
  \caption{ The strength of several modes in the disk as a function of
  time.  The solid line refers to the global disk eccentricity
  $S_{1,0}$. In the exponential growth regime the time derivative of
  the eccentricity (dotted line) is proportional to the (2,1) wave
  mode (short-dashed line).  }
   \label{fig:res}
\end{figure}

\subsection{Theoretical analysis}
The observed growth of the disk eccentricity in our simulations resembles
that found by \citet{2001A&A...366..263P} for very massive planets
with $M_{p} \gsim 10 M_{Jup}$.  The effect can be explained by a tidally
driven eccentricity through resonant interaction of the disk with
particular components of the planet's gravitational potential
\citep{1991ApJ...381..259L}. Using cylindrical coordinates $(r,
\varphi)$ we decompose the potential of the planet, which is on a {\it
circular orbit}, in the form 
\beq
\label{eq:potent}
     \Phi_p (r, \varphi) = \sum_{m=0}^{m=\infty}  \phi_m(r)
     \cos [ m ( \varphi - \Omega_p t) ]
\eeq
where $\Omega_p$ is the angular frequency of the planet.
The response of the disk has the form
\[
          \propto \exp [ i ( k \varphi - l \Omega_p t) ]
\]
The planetary potential produces tides in the disk, which interact
with an initially small eccentric disk.  The $m$-th Fourier component
of the potential (in Eq.~\ref{eq:potent}) excites an eccentric
Lindblad resonance in the {\it outer disk} where the rotation period
of the disk is $\Omega = {m\over{m+2}} \Omega_P$ which corresponds to
the mode $(k,l) = (m+1,m)$ \citep{1991ApJ...381..259L}.  Hence, for an
eccentric ($m=1$) perturbation the radial location lies at the outer
1:3 resonance at $r = 2.08$.  As the mass of the planetary companion
increases, the gap it opens in the disk will be deeper and
wider. Already in \citet{1992PASP..104..769A} it was suggested that
for sufficiently wide gaps, eccentricity growth can be induced by
interaction at the 1:3 resonance in the outer disk, However, for
smaller planet masses this is damped by other resonances which are
listed in \citet{2003ApJ...585.1024G, 2004ApJ...606L..77S}. The main
contributing eccentricity-damping resonances are the co-orbital
resonances and the resonances located at the outer 1:2 resonance.
Only if the gap is deep and wide enough these two resonances can no
longer cancel the eccentricity-exciting effect of the interaction at
the 1:3 resonance.  The radial surface density profiles for
simulations with different planet masses at 2500 orbits have been
displayed in Fig.~\ref{fig:sig1-mp}.  As can be seen, only for planet
masses larger than approximately 3 $M_{Jup}$ the gap is sufficiently
cleared at the 1:2 resonance ($r \approx$ 1.58) to allow for an
eccentricity increase of the disk.
 
Theoretical analysis in \citet{1991ApJ...381..268L} defines the total mode
strength $S_{k,l}$ as
\[
S_{k,l} = (S^2_{cos,cos,k,l} + S^2_{cos,sin,k,l} 
    + S^2_{sin,cos,k,l} + S^2_{sin,sin,k,l})^{1/2}
\] with $S_{f,g,k,l}$, defined in the inertial frame, given by

\begin{eqnarray*}
S_{f,g,k,l}& = & {2\over{\pi M (1+\delta_{i,0})(1+\delta_{j,0})}}\int\limits_t^{t+2\pi}dt'\int dr \int\limits_0^{2\pi} r d\theta \\ & &  \times \Sigma (r,\theta,t) f(i\theta)g(jt')
\end{eqnarray*}

In his analysis it is shown that the time derivative of the
 $k=1, l=0$ mode $S_{1,0}$ is given by the 
$k=2, l=1$ component:

\begin{equation}
\label{eq:proport}
         \frac{ S_{1,0}}{dt}  \propto  S_{2,1} \cdot S_{1,1}
\end{equation}

The evolution of the relevant mode strengths for a model with
$q=0.005$ is displayed in Fig.~\ref{fig:res}.  The amplitude of the
global eccentric mode ($k=1, l=0$) shows exponential growth (see
Fig.~\ref{fig:res}, solid line).  Furthermore, Eq.~(\ref{eq:proport})
is confirmed directly by comparing the $S_{2,1}$-mode (short-dashed
line) and the numerically obtained derivative of the eccentricity,
i.e. $S_{1,0}$ (dotted curve).  As it can be seen from the plot,
$S_{1,1}$ is constant as suggested by the theoretical analysis of
\citet{1991ApJ...381..259L}.

The good agreement of our results with theoretical expectations
supports our conclusion that the mechanism for eccentricity growth is
that described by \citet{1991ApJ...381..259L} and
\citet{2001A&A...366..263P}.  In our simulations growth will start
after the disk has settled sufficiently and the gap has been cleared,
a process which occurs on viscous time scales.  Our numerical growth
rates during the eccentricity increase have been estimated from the
time evolution of total radial kinetic energy
(Fig.~\ref{fig:ekingrow}).
\section{Conclusions}
We have performed numerical time dependent hydrodynamical calculations
of embedded planets in viscous accretion disks.  During the evolution
the planet is held fixed on a circular orbit, and the whole system is
evolved in time until a quasi-equilibrium state has been reached.  In
contrast to previous existing simulations on this problem we have
extended the evolutionary time to several thousand orbits of the
embedded planet for a whole range of different planetary masses.

We find that beyond a certain critical mass of the planet the
structure of the disk changes from a circular to an eccentric
state.  For typical viscosities in protoplanetary disks $\nu =
10^{-5}$ (or $\alpha \approx 0.004$) the transition to the eccentric
case occurs already for critical masses of $M_{p} = 3 M_{Jup}$.
Through a modal analysis we demonstrate that the eccentric ($k=1,
l=0$) mode in the disk is indeed driven by the ($k=2, l=1$) wave mode
which is excited at the outer 1:3 Lindblad resonance.  The numerically
inferred growth rate of the unstable eccentric disk mode is roughly
proportional to ${M_p}^{n}$ with $n=2.4$, which is slightly larger
that the predicted value of $n=2.0$
\citep{1991ApJ...381..259L,2001A&A...366..263P}. The discrepancy is
most likely due to a change in the density structure of the gap for
different planetary masses.  For small masses $M_{p} = 2 M_{Jup}$ no
eccentricity growth has been found. Here the damping effects of disk
viscosity and pressure keep the disk in the circular state.
Upon increasing the planetary mass the eccentricity eventually
saturates at a value of $e \approx 0.25$.

The excitation of eccentric disk modes by massive companions has been
studied within the framework of Cataclysmic Variable stars as an
instability of the inner disk
\citep{1991ApJ...381..259L,1991ApJ...381..268L}. In those cases the
change in viscous dissipation induced by the slow precession of the
disk is presently the preferred mechanism to explain the observed
superhumps in the light curve of some systems.  That the same process
is also applicable to (outer) disks around an embedded protoplanet has
been confirmed by \citet{2001A&A...366..263P} in their study of very
massive planets.
In their simulations a much larger threshold mass 
($\approx 10-20 \Mjup$) has been found. However, their
simulations were run only for 800 planetary orbits or less,
which is not sufficient to see growth for small mass planets
considering the long growth time of
the eccentric mode.

The change in the state of the disk has significant consequences for
the mass accretion rate onto the planet.  For circular disks the
width of the gap widens upon an increase in the planetary mass which
shuts off eventually further accretion of disk material. The maximum
mass a planet may reach by this process is around $5 \Mjup$
\citep{1999ApJ...514..344B, 1999ApJ...526.1001L}.  We suggest that
through the excitation of the eccentric mode in the disk the planet
can reach larger masses more readily, as there are quite a few systems
with (minimum) planetary masses larger than $5 M_{Jup}$.  The
influence an eccentric disk might have on the evolution of a pair of
planets engaged in a 2:1 resonance has been analyzed recently by
\citet{2005A&A...437..727K}. Here, changes in the libration amplitude
of the resonant angles are to be expected.

It has been suggested that the gravitational back reaction of an
embedded planet with the surrounding disk can lead to an increase in
the orbital eccentricity of the planet \citep{2003ApJ...585.1024G,
2003ApJ...587..398O}, and may serve as a possible mechanism to explain
the observed high eccentricities in extrasolar planetary systems.  In
the present work the gravitational back reaction of such an eccentric
disk on the planetary orbit has not been analyzed, and remains to be
studied in the future.  The magnitude of the reachable eccentricity
depends on the absolute physical mass of the ambient disk.  Through
numerical simulations \citet{2001A&A...366..263P} find that a
significant increase in planetary eccentricity is only seen for a
planet mass above $10 \, M_{Jup}$.  However, even in this case the
maximum eccentricities do not increase beyond $e=0.25$. Additionally,
the evolution time of the models was very short and did not allow to
study the longterm evolution of the eccentricity.

As the effect of disk eccentricity scales with planet mass at least as
$\propto {M_{p}}^{2.4}$ the effect is most pronounced for very massive
planets. However, in that case it is also more difficult to induce
high planetary eccentricities.  Hence, it is very questionable if the
back reaction of the disk can produce the observed high eccentricities
found in the surveys.

The present study is only two dimensional and has not included any
thermal effects such as radiative cooling or transport.  Since in two
dimensional calculations the gravitational effect between planet and
disk tends to be over-estimated (as the disk is confined to the
equatorial plane) one might expect a reduced effect in full
three-dimensional simulations. But the very low value of the critical
transition mass leaves sufficient room for an importance of this
effect in the growth of extrasolar planets.
\begin{acknowledgements}
We would like to thank Stephen Lubow, Doug Lin and Richard Nelson
for stimulating discussions during the course of this project.
The work was sponsored by the EC-RTN Network {\it The Origin of Planetary Systems}
under grant HPRN-CT-2002-00308.
\end{acknowledgements}
\bibliographystyle{aa}
\bibliography{3914}

\begin{thebibliography}{24}
\expandafter\ifx\csname natexlab\endcsname\relax\def\natexlab#1{#1}\fi

\bibitem[{{Artymowicz}(1992)}]{1992PASP..104..769A}
{Artymowicz}, P. 1992, \pasp, 104, 769

\bibitem[{{Bate} {et~al.}(2003){Bate}, {Lubow}, {Ogilvie}, \&
  {Miller}}]{2003MNRAS.341..213B}
{Bate}, M.~R., {Lubow}, S.~H., {Ogilvie}, G.~I., \& {Miller}, K.~A. 2003,
  \mnras, 341, 213

\bibitem[{{Bryden} {et~al.}(1999){Bryden}, {Chen}, {Lin}, {Nelson}, \&
  {Papaloizou}}]{1999ApJ...514..344B}
{Bryden}, G., {Chen}, X., {Lin}, D.~N.~C., {Nelson}, R.~P., \& {Papaloizou},
  J.~C.~B. 1999, \apj, 514, 344

\bibitem[{{Bryden} {et~al.}(2000){Bryden}, {R{\' o}{\. z}yczka}, {Lin}, \&
  {Bodenheimer}}]{2000ApJ...540.1091B}
{Bryden}, G., {R{\' o}{\. z}yczka}, M., {Lin}, D.~N.~C., \& {Bodenheimer}, P.
  2000, \apj, 540, 1091

\bibitem[{{D'Angelo} {et~al.}(2002){D'Angelo}, {Henning}, \&
  {Kley}}]{2002A&A...385..647D}
{D'Angelo}, G., {Henning}, T., \& {Kley}, W. 2002, \aap, 385, 647

\bibitem[{{Goldreich} \& {Sari}(2003)}]{2003ApJ...585.1024G}
{Goldreich}, P. \& {Sari}, R. 2003, \apj, 585, 1024

\bibitem[{{Goldreich} \& {Tremaine}(1980)}]{1980ApJ...241..425G}
{Goldreich}, P. \& {Tremaine}, S. 1980, \apj, 241, 425

\bibitem[{{Kley}(1989)}]{1989A&A...208...98K}
{Kley}, W. 1989, \aap, 208, 98

\bibitem[{{Kley}(1998)}]{1998A&A...338L..37K}
---. 1998, \aap, 338, L37

\bibitem[{{Kley}(1999)}]{1999MNRAS.303..696K}
---. 1999, \mnras, 303, 696

\bibitem[{{Kley} {et~al.}(2001){Kley}, {D'Angelo}, \&
  {Henning}}]{2001ApJ...547..457K}
{Kley}, W., {D'Angelo}, G., \& {Henning}, T. 2001, \apj, 547, 457

\bibitem[{{Kley} {et~al.}(2005){Kley}, {Lee}, {Murray}, \&
  {Peale}}]{2005A&A...437..727K}
{Kley}, W., {Lee}, M.~H., {Murray}, N., \& {Peale}, S.~J. 2005, \aap, 437, 727

\bibitem[{{Lubow}(1991{\natexlab{a}})}]{1991ApJ...381..259L}
{Lubow}, S.~H. 1991{\natexlab{a}}, \apj, 381, 259

\bibitem[{{Lubow}(1991{\natexlab{b}})}]{1991ApJ...381..268L}
---. 1991{\natexlab{b}}, \apj, 381, 268

\bibitem[{{Lubow} {et~al.}(1999){Lubow}, {Seibert}, \&
  {Artymowicz}}]{1999ApJ...526.1001L}
{Lubow}, S.~H., {Seibert}, M., \& {Artymowicz}, P. 1999, \apj, 526, 1001

\bibitem[{{Marcy} {et~al.}(2005){Marcy}, {Butler}, {Fischer}, {Vogt}, {Wright},
  {Tinney}, \& {Jones}}]{2005PThPS.158...24M}
{Marcy}, G., {Butler}, R.~P., {Fischer}, D., {et~al.} 2005, Progress of
  Theoretical Physics Supplement, 158, 24

\bibitem[{{Nelson} {et~al.}(2000){Nelson}, {Papaloizou}, {Masset}, \&
  {Kley}}]{2000MNRAS.318...18N}
{Nelson}, R.~P., {Papaloizou}, J.~C.~B., {Masset}, F.~S., \& {Kley}, W. 2000,
  \mnras, 318, 18

\bibitem[{{Ogilvie} \& {Lubow}(2003)}]{2003ApJ...587..398O}
{Ogilvie}, G.~I. \& {Lubow}, S.~H. 2003, \apj, 587, 398

\bibitem[{{Papaloizou} \& {Lin}(1984)}]{1984ApJ...285..818P}
{Papaloizou}, J. \& {Lin}, D.~N.~C. 1984, \apj, 285, 818

\bibitem[{{Papaloizou} {et~al.}(2001){Papaloizou}, {Nelson}, \&
  {Masset}}]{2001A&A...366..263P}
{Papaloizou}, J.~C.~B., {Nelson}, R.~P., \& {Masset}, F. 2001, \aap, 366, 263

\bibitem[{{Sari} \& {Goldreich}(2004)}]{2004ApJ...606L..77S}
{Sari}, R. \& {Goldreich}, P. 2004, \apjl, 606, L77

\bibitem[{{van Leer}(1977)}]{1977JCoPh..23..276V}
{van Leer}, B. 1977, Journal of Computational Physics, 23, 276

\bibitem[{{Ward}(1986)}]{1986Icar...67..164W}
{Ward}, W.~R. 1986, Icarus, 67, 164

\bibitem[{{Ziegler}(1998)}]{1998CoPhC.109..111Z}
{Ziegler}, U. 1998, Computer Physics Communications, 109, 111

\end{thebibliography}
\end{document}